\documentclass[pra,10pt,twocolumn,superscriptaddress,showpacs]{revtex4-2}

\usepackage{epsfig}
\usepackage{amsmath}
\usepackage{graphicx}
\usepackage{graphics}
\usepackage{float}
\usepackage{bm}
\usepackage{dsfont}
\usepackage{mathtools}
\usepackage{amssymb}
\usepackage[usenames,dvipsnames]{color}
\usepackage{natbib}
\usepackage{epstopdf}
\usepackage{verbatim}
\usepackage{wrapfig}
\usepackage{soul}

\DeclareMathOperator{\Tr}{\text{Tr}}

\graphicspath{{Figure/}}

\newcommand{\bra}[1]{\left\langle #1\right|}
\newcommand{\ket}[1]{\left| #1\right\rangle}

\newcommand{\beq}{\begin{equation}}
\newcommand{\eeq}{\end{equation}}
\newcommand{\tr}{\text{Tr}}

\begin{document}

\title{The role of initial system-environment correlations with a spin environment}

\author{Ali Raza Mirza}
\affiliation{School of Science \& Engineering, Lahore University of Management Sciences (LUMS), Opposite Sector U, D.H.A, Lahore 54792, Pakistan}

\author{Mah Noor Jamil}

\affiliation{School of Science \& Engineering, Lahore University of Management Sciences (LUMS), Opposite Sector U, D.H.A, Lahore 54792, Pakistan}

\author{Adam Zaman Chaudhry}
\email{adam.zaman@lums.edu.pk}
\affiliation{School of Science \& Engineering, Lahore University of Management Sciences (LUMS), Opposite Sector U, D.H.A, Lahore 54792, Pakistan}

\begin{abstract}

Open quantum systems are a subject of immense interest as their understanding is crucial in the implementation of modern quantum technologies. In the study of their dynamics, the role of the  initial system-environment correlations is commonly ignored. In this work, to gain insights into the role of these correlations, we solve an exactly solvable model of a single two-level system interacting with a spin environment, with the initial system state prepared by a suitable unitary operation. By solving the dynamics exactly for arbitrary system-environment coupling strength while taking into account the initial system-environment correlations, we show that the effect of the initial correlations is, in general, very significant and non-trivial. To further highlight the importance of the initial system-environment correlations, we also extend our study to investigate the dynamics of the entanglement between two two-level systems interacting with a common spin environment. 

\end{abstract}


\maketitle

\section{Introduction}  

Quantum systems generally interact with their environment, thereby allowing the environment to act as a probe that can make indirect measurements of the quantum states of the system. This results in the process of decoherence, wherein a set of pure states are selected and other superpositions are destroyed. Considering how non-trivial such a system can be, many approximations and assumptions have to be employed in order to understand the dynamics  of such open quantum systems \cite{BPbook,Weissbook}. For example, to apply perturbation theory, one has to assume that the interaction of the system and environment is weak. Furthermore, to disregard any memory effects, one also has to assume a short environment correlation time. For weak system-environment interaction, initial system-environment correlations are ignored by considering an initial product state of the system and environment, with the environment in its thermal equilibrium state. Moreover, with a small environment correlation time, the environment tends to lose information about the system quickly, which further justifies ignoring the effect of the initial system-environment correlations \cite{Modi2011}. It must be emphasized that all such approximations fail to work as soon as we consider strong system-environment interaction strength.

With the rising interest in different practical quantum systems ranging from superconducting qubits and quantum dots to light-harvesting complexes involving strong system-environment interactions, many studies have been performed to better analyze the effects of the initial system-environment correlations \cite{HakimPRA1985, HaakePRA1985, Grabert1988, SmithPRA1990, GrabertPRE1997, PazPRA1997, LutzPRA2003, BanerjeePRE2003, vanKampen2004, BanPRA2009, HanggiPRL2009, UchiyamaPRA2010, TanimuraPRL2010, SmirnePRA2010, DajkaPRA2010, ZhangPRA2010,TanPRA2011, CKLeePRE2012,MorozovPRA2012, SeminPRA2012,  ChaudhryPRA2013a,ChaudhryPRA2013b,ChaudhryCJC2013,FanchiniSciRep2014,FanSciRep2015,ChenPRA2016,VegaRMP2017,VegaPRA2017,ShibataJPhysA2017,CaoPRA2017}. Unfortunately, the effect of the initial correlations are most pronounced in the strong system-environment coupling regime where the usual perturbative methods fail to work \cite{BPbook}. To counter this problem, one can study exactly solvable models - see, for example, Refs.~\cite{MorozovPRA2012} and \cite{ChaudhryPRA2013a}. However, the study of such initial correlations as done in Refs.~\cite{MorozovPRA2012} and \cite{ChaudhryPRA2013a} are limited in that they use exactly solvable dephasing models only, thereby rendering them unable to describe dissipative effects. Simply put, any changes in the diagonal elements of the system density matrix remain unaccounted for, and this misses a major piece in the puzzle of understanding the dynamics of the open quantum system upon interacting with the environment.

Our objective in this work is to study an exactly solvable model in which the initial system-environment correlations are included, thereby allowing for both dissipative and dephasing effects to be better analyzed. In this sense, we further extend a previously studied model of a single two-level system (or spin) interacting with an environment consisting of a collection of spins \cite{CucchiettiPRA2005}. The solution will be non-trivial, as the system spin Hamiltonian does not commute with the system-environment interaction. Once the system and the environment are allowed to reach a joint equilibrium state, a unitary operator is applied on the system only to prepare the desired initial state. Generally speaking, this joint system-environment state is correlated, making it starkly different from the usually used simplistic uncorrelated product states \cite{CKLeePRE2012}. The state preparation is expected to influence the dynamics of the system spin through the system-environment correlations present before the state preparation \cite{MorozovPRA2012,ChaudhryPRA2013a,ChaudhryPRA2013b,ShibataJPhysA2017}. The obvious benefit of such a model is that it provides simple expressions for the evolution of the Bloch vector of the system's spin for arbitrary temperature and system-environment coupling strength. Through exact analytical solutions for the Bloch vector, we can show that the state preparation heavily affects the system dynamics via the initial system-environment correlations. More specifically, we prove that, generally speaking, with lower temperatures and stronger system-environment coupling strengths, the initial correlations can significantly influence the subsequent dynamics; on the other hand, with higher temperature and weak system-environment correlations, the effect of the initial correlations is much smaller. We then extend our model to two spins interacting with a common spin environment to demonstrate the importance of the role of the initial correlations. Let us also note that a similar study has been performed before \cite{majeed2019effect}. However, that study considered the initial state preparation via a selective projective measurement while we are considering the more experimentally friendly scenario of the initial state preparation via a unitary operation. Moreover, with a projective measurement, the initial systsem-environment state is again a product state; however, the effect of the initial correlations are exhibited via the modified initial environment state. With a unitary operator preparing the initial state, the system and the environment remain, in general, correlated. 

We organize this paper as follows. In section \ref{spinspin}, we present our model, discuss the initial state preparation, and present results for the central two-level system dynamics, both with and without initial system-environment correlations. Thereafter, in section \ref{2qubits}, we consider two central spins coupled to the common environment and focus on the entanglement dynamics. We finally conclude in Sec.~IV.

\section{Spin-spin Model}

\label{spinspin}
We consider a single spin-half system (a qubit) interacting with $N$ spin-half systems (the spin environment). We write the system-environment Hamiltonian as 
\begin{align}
    H_{\text{tot}} =
    \begin{cases}
      H_{\text{S0}} + H_E + H_{\text{SE}} & t\le 0,\\
      H_{S} + H_E + H_{\text{SE}} & t > 0.\\
    \end{cases}   
\end{align}
Here $H_E$ is the Hamiltonian of the spin environment alone, and $H_{\text{SE}}$ is the system-environment interaction. We prepare a desired initial state at time $t = 0$; the system Hamiltonian after this state preparation process can be different as compared to the system Hamiltonian before the state preparation process. As such, $H_S$ denotes the system Hamiltonian corresponding to the coherent evolution of the system only after the initial time $t = 0$ at which the system state is prepared. $H_{\text{S0}}$ is the system Hamiltonian before the system state preparation, with the parameters in $H_{\text{S0}}$ chosen so as to aid the state preparation process. Note that $H_{\text{S0}}$ is similar to $H_S$ in the sense that both operators live in the same Hilbert space, but they may have different parameters. For the spin-spin model that we are discussing, we have (we take $\hbar = 1$ throughout this paper) 
\begin{align}
    H_{\text{S0}}
    &= \frac{\varepsilon_0}{2} \sigma_z +\frac{\Delta_0}{2} \sigma_x,
\\
 H_{S}
    &= \frac{\varepsilon}{2} \sigma_z +\frac{\Delta_0}{2} \sigma_x,
\\
    H_E
    &=\sum_{i=1}^{N} \frac{\varepsilon_i}{2} \sigma^{(i)}_z + \sum_{i=1}^{N} \alpha_i \sigma^{(i)}_z \sigma^{(i+1)}_z ,
\\
    H_{\text{SE}}
    &= \frac{1}{2} \sigma_z \otimes \sum_{i=1}^{N} g_i\sigma^{(i)}_z .
\end{align}
Here $\sigma_{i}$  $ (i = x, y, z) $ represent the usual Pauli spin matrices, $\varepsilon_0$ and $\varepsilon$ denote the energy-level spacing of the central spin before and after the state preparation respectively, $\Delta_0$ is the tunneling amplitude, and $\varepsilon_{i}$ denotes the energy level spacing for the $i^{\text{th}}$ environmental spin. 
We also allow environment spins to interact with each other via $\sum_{i=1}^{N} \alpha_i \sigma^{(i)}_z \sigma^{(i+1)}_z$, where $\alpha_i$ denotes the nearest neighbor interaction strength between the environment spins. The central spin interacts with the environment spins through $H_{\text{SE}}$, where $g_i$ is the interaction strength between the central qubit and the $i^{\text{th}}$ environment spin. Note that our system Hamiltonian $H_S$ commutes with the total Hamiltonian meaning that the system energy is conserved. 

Our primary goal is to find the dynamics of our central qubit system. To do that, we first obtain the total unitary time evolution operator $U(t)$ for the system and its environment as a whole. We write $H_{\text{SE}}=S\otimes E$, where $S$ is a system operator and $E$ is an environment operator. Now, the states $\ket{n}=\ket{n_1}\ket{n_2}\ket{n_3}...\ket{n_N} $ are the eigenstates of $E$ with $n_i=0$ signifying the spin-up along $z$ state while $n_i = 1$ is the spin-down state. We then have
\begin{align}
    E\ket{n}
    = e_n\ket{n},
\end{align}
with $e_n= \sum^N_{i=1} (-1)^{n_i} g_i$. We also have
\begin{align}
    \sum_{i=1}^{N} \varepsilon_i \sigma^{(i)}_z\ket{n}
    = \epsilon_n \ket{n},\label{eq1}
\\
    \sum_{i=1}^{N} \alpha_i\sigma^{(i)}_z\sigma^{(i+1)}_z  \ket{n}
    = \lambda_n \ket{n},\label{eq2}
\end{align}
where $\epsilon_n= \sum^N_{i=1} (-1)^{n_i} \varepsilon_i$ and $\lambda_n= \sum^N_{i=1}\alpha_i (-1)^{n_i}(-1)^{n_{i+1}}$ are the eigenvalues of the first and second terms of the environment Hamiltonian respectively. 

\subsection{\label{sec:level2A}Initial state preparation without correlations}
We now discuss the preparation of the initial system state. Ignoring the system-environment correlations, we can write the system-environment equilibrium state as a product state, namely 
\begin{align}
    \rho 
    = \rho_{\text{S0}} \otimes \rho_{E}.\label{eq6}
\end{align}
Here $\rho_{\text{S0}} =e^{-\beta H_{\text{S0}}}/Z_{\text{S0}}$ and $ \rho_{E} = e^{-\beta H_E}/Z_E$ with the partition functions $Z_{\text{S0}} = \tr_S \left\{e^{-\beta H_{\text{S0}}}\right\}$ and $Z_E = \tr_E \left\{e^{-\beta H_E}\right\}$. $\beta$ is the inverse of temperature with $k_{B}=1$. Note that writing the state in this form is only justified if we can ignore the system-environment coupling $H_{\text{SE}}$ (or, in other words, we are in the weak coupling regime) since $H_{\text{SE}}$ does not commute with the system Hamiltonian. Now, a relatively large value of $\varepsilon_0$ and a small value of $\Delta_0$, that is, $\beta \varepsilon_0 \gg 1$, will correspond to the system state being approximately `down' along the $z\text{-}$axis. At time $t = 0$, we apply a unitary operator to prepare the desired initial state. For example, if the desired initial state is `spin up' along the $x\text{-}$axis, then the unitary operator $R= e^{i \frac{\pi}{4} \sigma_y}$, realized by the application of a suitable control pulse, is applied to the system only. During the pulse operation, we assume that the pulse duration is much smaller than the cutoff frequency of the environment $\omega_c$ and the effective Rabi frequency  $\sqrt{\varepsilon^2 + \Delta^2}$. Once the pulse has been applied, the total system-environment initial state is
\begin{equation}
    \rho^R_{\text{tot}}
     = \rho^R_{\text{S0}} \otimes \rho_{E}, \label{eq3}
\end{equation}
with $ \rho^R_{\text{S0}} =  e^{-\beta H^R_{\text{S0}}} /Z_{\text{S0}} $ and $H^R_{\text{S0}} = R H_{\text{S0}} R^{\dagger}$. Once we have prepared our system's initial state, we can change the system Hamiltonian parameters as needed. For example, we can change the energy bias to $\varepsilon$ so that the contribution of the tunneling term ($\frac{\Delta}{2} \sigma_x$) becomes more significant. Once again, we assume that this change occurs in a very short time duration. We now write the initial system state as (the superscript `woc' stands for `without correlations' since we are ignoring the system-environment interaction when preparing the initial system state) 
\begin{align}
    \rho_{\text{S0}}^{\text{woc}}
    =\frac{1}{Z_{\text{S0}}} \left\{\mathds{1} \cosh	\left(\beta\widetilde{\Delta}_0\right) - \frac{\sinh\left(\beta\widetilde{\Delta}_0\right)}{\widetilde{\Delta}_0}H^{R}_{\text{S0}}\right\},\nonumber
\end{align}
with $\widetilde{\Delta}_0 = \frac{1}{2} \sqrt{\varepsilon_0^2 + \Delta_0^2} $. It is convenient to find the Bloch vector components using $ p^{\text{woc}}_i = \tr_S \left\{\sigma_i  \rho_{\text{S0}}^{\text{wc}} \right\} $ (with $i= x,y,z$) and cast them into the column vector
\begin{equation}
    \left( \begin{array}{c}
   p^{\text{woc}}_x\\
   p^{\text{woc}}_y\\
   p^{\text{woc}}_z
  \end{array} \right)
  	= \frac{\sinh\left(\beta\widetilde{\Delta}_0\right)}{Z_{\text{S0}}\widetilde{\Delta}_0}
	\left( {\begin{array}{c}
  	\varepsilon_0\\
   	0\\
   -\Delta_0\\
  \end{array} } \right). \label{eq10}
\end{equation}

\subsection{\label{sec:level2A}Initial state preparation with correlations}

We now consider the initial system-environment state that includes the effect of the initial system-environment correlations. We imagine that the spin system has been interacting with its surrounding environment for very a long time before coming to a joint thermal equilibrium state with the environment; the system-environment state is then the standard canonical Gibbs state $\rho_{\text{th}}= e^{-\beta H}/Z_{\text{tot}}$. In general, we can not write this state as a product state since the system-environment interaction does not commute with the system Hamiltonian. However, it is quite clear that if the system-environment coupling is weak, this state would approximate the product state given in Eq.~\eqref{eq6}. Now, at time $t=0$, as before, we apply a suitable pulse to prepare the initial system state. Consequently, the correlated system-environment state becomes (the superscript `wc' stands for `with correlations')
\begin{align}
\rho^{\text{wc}}_{\text{tot}}
&= \frac{1}{Z_{\text{tot}}} e^{-\beta \left(H^R_{\text{S0}} + H_E + H^R_{\text{SE}}\right)},\label{eq5}
\end{align}
where $Z_{\text{tot}} = \tr_{\text{SE}} \left\{ e^{-\beta \left(H^R_{\text{S0}} + H_E + H^R_{\text{SE}}\right)} \right\}$ is the combined partition function for the system and the environment as a whole. Looking at Equations \eqref{eq1} and \eqref{eq2}, we can write $e^{-\beta H_E} \ket{n}
    = k_n \ket{n}$ with $k_n = e^{-\beta (\frac{\epsilon_n}{2} + \lambda_n)}$. Also,
\begin{align}
    \left(H^R_{\text{S0}} + H^R_{\text{SE}}\right)\ket{n}
    &= \left(\frac{\varepsilon^n_{0}}{2} \sigma_z  -\frac{\Delta_0}{2} \sigma_x\right)\ket{n} \equiv H_{\text{S0},n}\ket{n},\nonumber
\end{align}
where $H_{\text{S0},n}$ is a `shifted' system Hamiltonian due to the system-environment interaction with the new parameter $\varepsilon_{0,n} = e_n + \varepsilon_0$. Following the same steps as in the previous section, we can eventually write
\begin{equation}
    \left( \begin{array}{c}
   p^{\text{wc}}_x\\
   p^{\text{wc}}_y\\
   p^{\text{wc}}_z
  \end{array} \right)
  	= \sum_n \frac{k_n\sinh\left(\beta\widetilde{\Delta}^n_0\right)}{Z_{\text{tot}}\widetilde{\Delta}^n_0}
	\left( {\begin{array}{c}
  	\varepsilon^n_0\\
   	0\\
   -\Delta_0\\
  \end{array} } \right),    
\end{equation}
where we now have $\widetilde{\Delta}^n_0 = \frac{1}{2}\sqrt{(\varepsilon_0^{n})^2 + \Delta_0^2} $.

\subsection{\label{sec:level2A}System dynamics without initial correlations}

To find the dynamics, we construct the total time-evolution unitary operator by inserting the completeness relation over the environment states $\ket{n}$, that is, over all the possible environment spin orientations. This gives us
\begin{align}
    U(t)
    &=\sum_{n} e^{-i\frac{\epsilon_n}2t}e^{-i\lambda_nt}e^{-iH_{S,n}t}\ket{n}\bra{n},\nonumber
\\
    &=\sum^{2^{N}-1}_{n=0} U_n(t)\ket{n}\bra{n},\label{eq4}
\end{align}
where $H_{S,n}$ is similar to $H_{\text{S0},n}$, the only difference being that the latter contains the energy bias $\varepsilon_{0,n}$ and the former $\varepsilon_n = e_n + \varepsilon$. Now, we can write
\begin{align}
    U_n(t)
    &=e^{-i\frac{\epsilon_n}2t}e^{-i\lambda_nt}\left\{\mathds{1}\cos\left(\widetilde{\Delta}_nt\right)-\frac{i\sin\left(\widetilde{\Delta}_nt\right)}{\widetilde{\Delta}_n} H_{S,n} \right\},
\end{align}
which is the effective unitary operator that only acts in the system's Hilbert space, with $\widetilde{\Delta}_n = \sqrt{\varepsilon^2 + \Delta^2} $. The reduced density matrix for the system at time $t$ can then be obtained via $\rho^{\text{woc}}_S(t)
    =\tr_E\left\{U(t)\rho^{R}_{\text{tot}}U^\dagger(t)\right\}$. Upon taking $U(t)$ from \eqref{eq4}, and the simple product state $\rho^{R}_{\text{tot}}$ from \eqref{eq3}, we obtain, after some algebra, 
\begin{align}
    \rho^{\text{woc}}_S(t)
    &=\frac{1}{Z_E}\sum^{2^{N}-1}_{n=0}k_n{U_n(t)\rho_{\text{S0}}^{\text{woc}}U^\dagger_n(t)}.
\end{align}
Here $Z_{E}=\sum_n k_n$ which is sensible because every environment spin configuration $\ket{n}$ occurs with probability $k_n/Z_E$. $U_n(t)$ generates dynamics for each configuration, meaning that to obtain the total reduced density matrix for the system, we need to take into account all the possible environment spin configurations. 

It is useful to find the Bloch vector components for the time-evolved density matrix. We can determine the Bloch vector $\mathbf{p}(t)$ at time $t$ via $\mathbf{p}^{\text{woc}}(t) = \frac{1}{Z_E}\mathbf{M}^{\text{woc}}(t)\mathbf{p}^{\text{woc}}$. Written out explicitly, this is  
\begin{multline}
\left( {\begin{array}{c}
   p^{\text{woc}}_x(t)\\
   p^{\text{woc}}_y(t)\\
   p^{\text{woc}}_z(t)
  \end{array} } \right) = \frac{1}{Z_E} \left( {\begin{array}{ccc}
   M^{\text{woc}}_{11} & M^{\text{woc}}_{12} & M^{\text{woc}}_{13} \\
   M^{\text{woc}}_{21} & M^{\text{woc}}_{22} & M^{\text{woc}}_{23} \\
   M^{\text{woc}}_{31} & M^{\text{woc}}_{32} & M^{\text{woc}}_{33}
  \end{array} } \right) \left( {\begin{array}{c}
   p^{\text{woc}}_x\\
   p^{\text{woc}}_y\\
   p^{\text{woc}}_z
  \end{array} } \right),
  \end{multline}
with 
\begin{align}
M^{\text{woc}}_{11}(t) 
&= \sum_{n} \frac{k_n}{4\widetilde{\Delta}_n^2} \left[\Delta^2 + \varepsilon_n^2 \cos(2\widetilde{\Delta}_n t) \right], \notag 
\\
M^{\text{woc}}_{12}(t) 
&= -\sum_{n} \frac{k_n \varepsilon_n}{2\widetilde{\Delta}_n}  \sin(2\widetilde{\Delta}_n t), \notag
\\
M^{\text{woc}}_{13}(t) 
&= \sum_{n} \frac{k_n \Delta \varepsilon_n}{2\widetilde{\Delta}_n^2} \sin^2(\widetilde{\Delta}_n t),\notag
\\
M^{\text{woc}}_{21}(t) 
&= \sum_{n} \frac{k_n \varepsilon_n}{2\widetilde{\Delta}_n}  \sin(2\widetilde{\Delta}_n t), \notag
\\
M^{\text{woc}}_{22}(t) 
&= \sum_{n} k_n \cos(2\widetilde{\Delta}_n t), \notag
\\
M^{\text{woc}}_{23}(t) 
&= -\sum_{n} \frac{k_n \Delta}{2\widetilde{\Delta}_n} \sin(2\widetilde{\Delta}_n t),\notag
\\
M^{\text{woc}}_{31}(t) 
&= \sum_{n} \frac{k_n\Delta \varepsilon_n}{2\widetilde{\Delta}_n^2}  \sin^2(\widetilde{\Delta}_n t),\notag 
\\
M^{\text{woc}}_{32}(t) 
&= \sum_{n} \frac{k_n \Delta}{2\widetilde{\Delta}_n} \sin(2\widetilde{\Delta}_n t),\notag 
\\
M^{\text{woc}}_{33}(t) 
&= \sum_{n} \frac{k_n}{4\widetilde{\Delta}_n^2} \left[\varepsilon_n^2 + \Delta^2 \cos(2\widetilde{\Delta}_n t)\right].
\end{align}
To find these matrix elements, we need to calculate sums over all possible $2^N$ environment configurations. We emphasize that this is an exact solution that is also valid even if the couplings $g_i$ was large. Furthermore, it is obvious that, in general, both the off-diagonal and diagonal elements of the system density matrix evolve with time.

\begin{figure}[t]
 		\includegraphics[scale = 0.8]{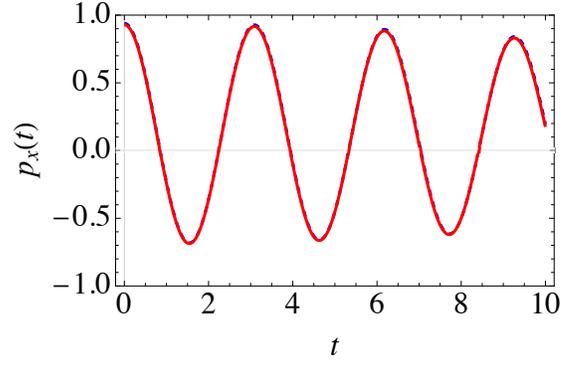}
 		\centering
		\caption{(Color online) Dynamics of $p_x(t)$ for relatively weak system-environment coupling without initial correlations (dashed, blue line) and with initial correlations (solid, red line). We work in dimensionless units throughout and we have set $\Delta_0 = 1$. Other system-environment parameters are $g_i = 0.01$, $\varepsilon_0 = 4, \varepsilon = 2$, $\varepsilon_i = 1$, $\beta = 1$ and $N = 50$.}
		\label{weakcoupling}
\end{figure}
\begin{figure}[t]
				\includegraphics[scale = 0.8]{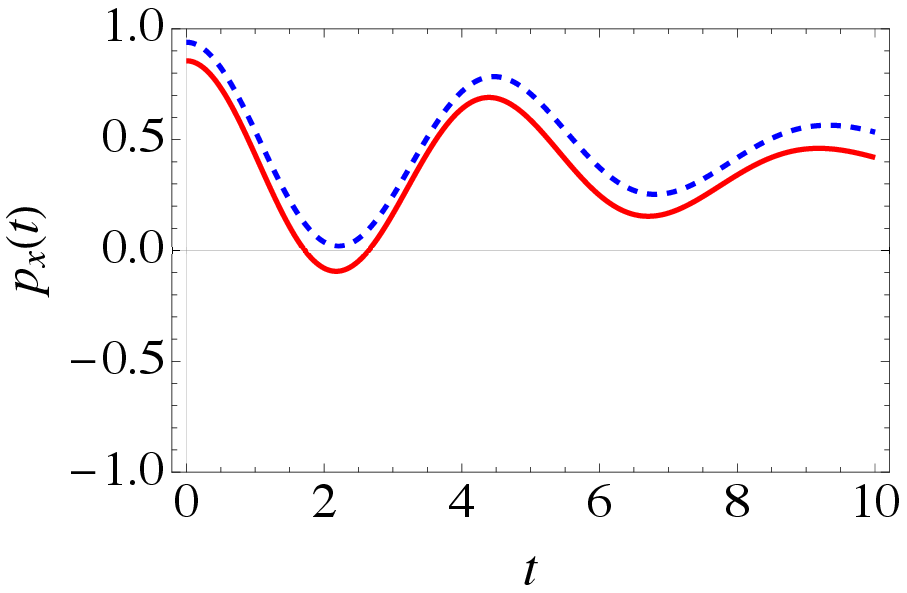}
 				\centering	
				\caption{(Color online) Same as Fig.~\ref{weakcoupling}, except that now we have  $g_i = 0.05$.}
				\label{midcoupling}
\end{figure}
\begin{figure}[t]
				\includegraphics[scale = 0.8]{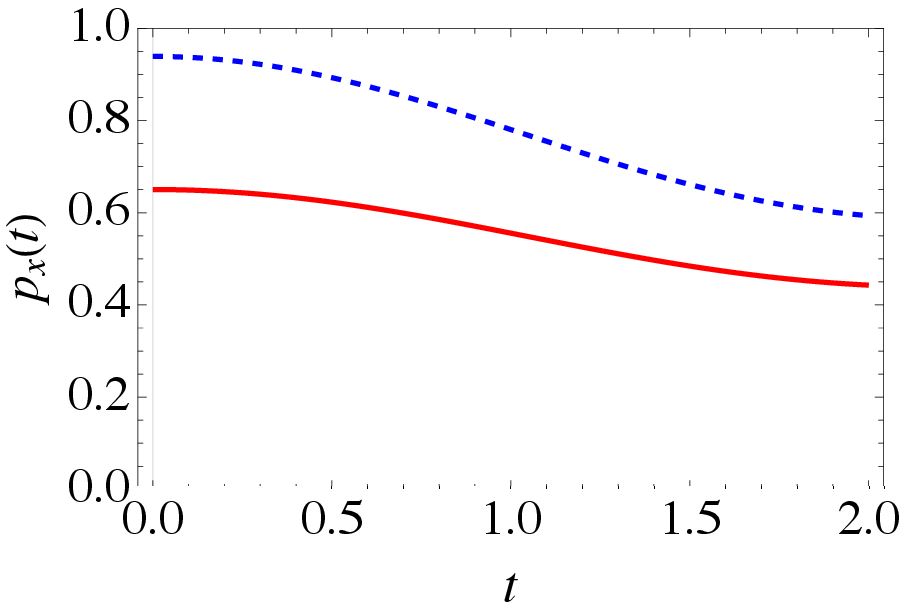}
 				\centering	
				\caption{(Color online) Same as Fig.~\ref{weakcoupling}, except that now we have  $g_i = 0.1$.}
				\label{comparison}
\end{figure}
\begin{figure}[t]
				\includegraphics[scale = 0.8]{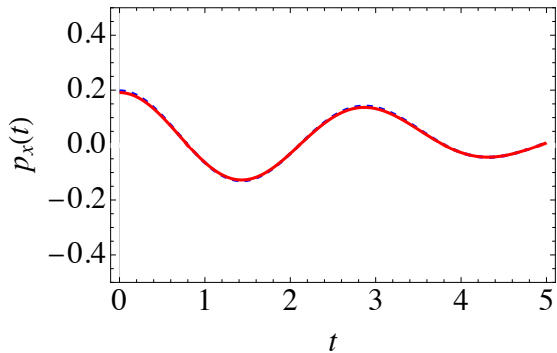}
 				\centering	
				\caption{(Color online) Same as Fig.~\ref{midcoupling}, except that now we have $\beta = 0.1$.}
				\label{hightemp}
\end{figure}

\subsection{Dynamics with correlated initial state}

We now study the dynamics while incorporating the initial system-environment correlations. We take the initial state given in Eq.~\eqref{eq5}, let it evolve under the unitary operator \eqref{eq4}, thereby taking a trace over the environment to obtain the following reduced system density matrix   
\begin{align}
    \rho^{\text{wc}}_S (t)=\frac{1}{Z_{\text{tot}}}\sum_{n}A_n k_n {U_n(t)\rho_{S}^{\text{wc}}U^\dagger_n(t)},
\end{align}
where now $Z_{\text{tot}}=\sum_n{A_n k_n}$ with $A_n = \tr\left\{e^{-\beta \left(\frac{\varepsilon^n_0}{2}\sigma_z + \frac{\Delta_0}{2}\sigma_x \right)}\right\} = 2\cosh\left(\beta \widetilde{\Delta}^n_0\right)$. The Bloch vector $\mathbf{p}(t)$ at time $t$ is now given by $\mathbf{p^{\text{wc}}}(t) = \frac{1}{Z_{\text{tot}}}\mathbf{M}^{\text{wc}}(t)\mathbf{p}^{\text{wc}}$, with 
\begin{align}
M^{\text{wc}}_{11}(t) 
&= \sum_{n} \frac{A_nk_n}{4\widetilde{\Delta}_n^2} \left[\Delta^2 + \varepsilon_n^2 \cos(2\widetilde{\Delta}_n t) \right], \notag 
\\
M^{\text{wc}}_{12}(t) 
&= -\sum_{n} \frac{A_nk_n \varepsilon_n}{2\widetilde{\Delta}_n}  \sin(2\widetilde{\Delta}_n t), \notag
\\
M^{\text{wc}}_{13}(t) 
&= \sum_{n} \frac{A_nk_n \Delta \varepsilon_n}{2\widetilde{\Delta}_n^2} \sin^2(\widetilde{\Delta}_n t),\notag
\\
M^{\text{wc}}_{21}(t) 
&= \sum_{n} \frac{A_nk_n \varepsilon_n}{2\widetilde{\Delta}_n}  \sin(2\widetilde{\Delta}_n t), \notag
\\
M^{\text{wc}}_{22}(t) 
&= \sum_{n} A_nk_n \cos(2\widetilde{\Delta}_n t), \notag
\\
M^{\text{wc}}_{23}(t) 
&= -\sum_{n} \frac{A_nk_n \Delta}{2\widetilde{\Delta}_n} \sin(2\widetilde{\Delta}_n t),\notag
\\
M^{\text{wc}}_{31}(t) 
&= \sum_{n} \frac{A_nk_n\Delta \varepsilon_n}{2\widetilde{\Delta}_n^2}  \sin^2(\widetilde{\Delta}_n t),\notag 
\\
M^{\text{wc}}_{32}(t) 
&= \sum_{n} \frac{A_nk_n \Delta}{2\widetilde{\Delta}_n} \sin(2\widetilde{\Delta}_n t),\notag 
\\
M^{\text{wc}}_{33}(t) 
&= \sum_{n} \frac{A_nk_n}{4\widetilde{\Delta}_n^2} \left[\varepsilon_n^2 + \Delta^2 \cos(2\widetilde{\Delta}_n t)\right].
\end{align}

Note that, once again, this is a non-perturbative solution. Comparing the time evolution of the system with the two different initial states, it is clear that the difference in the dynamics is due to the factor $A_n$ which encapsulates the effects of initial correlations before the state preparation. If these correlations are included, every possible environment configuration occurs with the probability $A_nk_n /Z_{\text{tot}}$ instead of $k_n/Z_E$, thus leading to a possibly marked difference in the evolution of the Bloch vector components. To examine this difference in more detail, let us note that as long as the system-environment coupling strength is weak, we expect negligible evolution differences between the dynamics of the correlated and uncorrelated initial state. As we increase the coupling strength, the effect of the initial correlations should look more prominent. These two forecasts are presented in Figs.~\ref{weakcoupling} and ~\ref{midcoupling}, where we have plotted $p^{\text{wc}}_x(t)$ (the $x\text{-}$component of the Bloch vector, taking the initial correlations into account) and $p^{\text{woc}}_x(t)$ (the $x\text{-}$component of the Bloch vector starting from the simple product state) as a function of time. Two points should be noted. First, the correlation effect is more pronounced in Fig.~\ref{midcoupling} (coupling strength $g=0.05$) as compared to Fig.~\ref{weakcoupling} where coupling strength is $ g = 0.01$. Second, as expected, with a stronger system-environment coupling, the oscillations in the Bloch vector dynamics die off more quickly. As the coupling strength is increased, the effect of the initial correlations becomes even more pronounced [see Fig.~\ref{comparison}]. 

\begin{figure}[t]
 		\includegraphics[scale = 0.8]{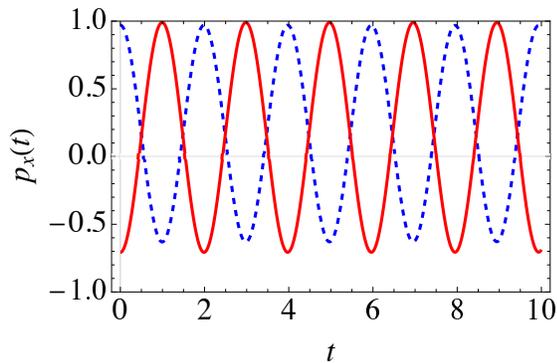}
 		\centering
		\caption{(Color online) Same as Fig.~\ref{weakcoupling}, except that now we have $g = 1$ and $\beta = 10$.}
		\label{lowtemp}
\end{figure}

We can also investigate the effect of temperature. At higher temperatures, the total system-environment thermal equilibrium state (before applying the pulse) is almost a mixed state. Hence, at higher temperatures, there will be little difference as both initial states are effectively the same. We illustrate this in Figs.~\ref{hightemp} and \ref{lowtemp}. There are two points to be made regarding Fig.~\ref{hightemp}. First, at higher temperatures, the condition $\beta\varepsilon_0 \gg 1$ is not fulfilled. Therefore, the system Bloch vector, before the pulse operation, is not approximately along the negative $z\text{-}$axis. Consequently, the evolution of the Bloch vector component $p_x(t)$ does not start from $p_x \approx 1 $. Second, the correlation effect seen in Fig.~\ref{midcoupling} disappears at higher temperatures although the coupling strength is still $g=0.05$ in Fig.~\ref{hightemp}. The dynamics with the two different initial states at even lower temperatures is illustrated in Fig.~\ref{lowtemp} where $\beta = 10$. For the simple product initial state, since $\beta\varepsilon_0 \gg 1$, the initial system state just after the pulse is applied is approximately $p_x=1$. If we instead consider the joint system-environment thermal equilibrium state, the interaction Hamiltonian term $H_{\text{SE}}$ dominates; this leads to the system state being approximately `up' along the $z\text{-}$axis before the application of the pulse and `down' along the $x\text{-}$axis after the pulse operation.

\begin{figure}[t]
		\includegraphics[scale = 0.8]{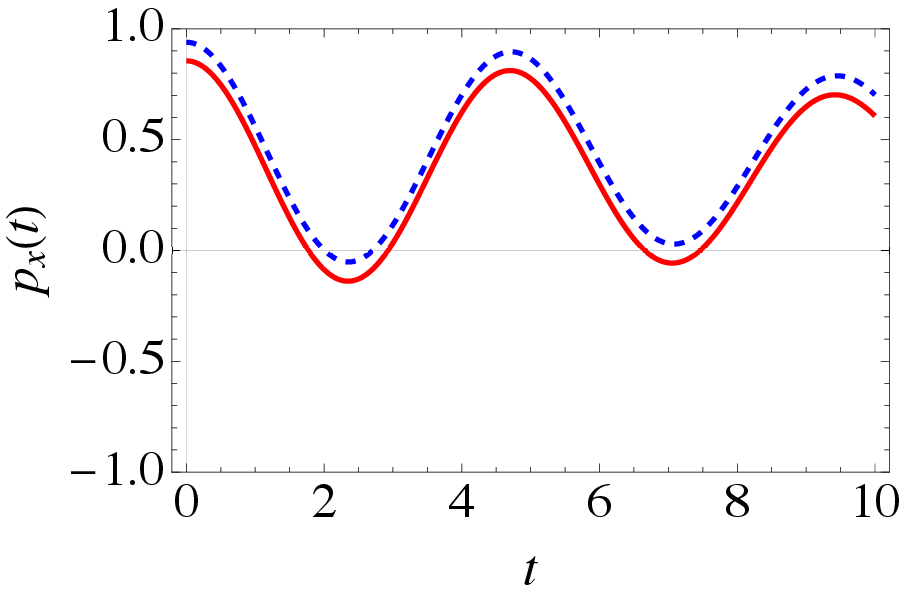}
 		\centering	
		\caption{(Color online) Same as Fig.~\ref{weakcoupling}, except that now we have $N = 250$.}
		\label{environment}
\end{figure}

\begin{figure}[t]
 		\includegraphics[scale = 0.8]{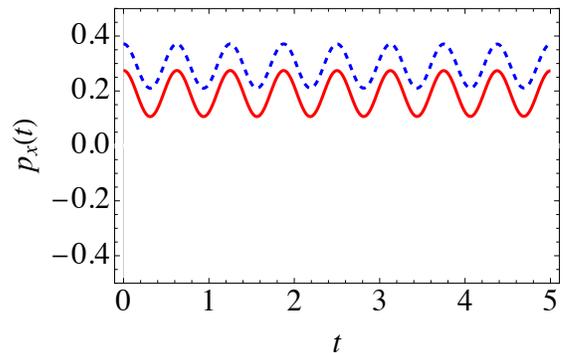}
 		\centering
		\caption{(Color online) Same as Fig.~\ref{weakcoupling}, except that now we have $\Delta_0 = 10$ and $g = 0.05$.}
		\label{Delta}
\end{figure}

\begin{figure}[t]
 		\includegraphics[scale = 0.8]{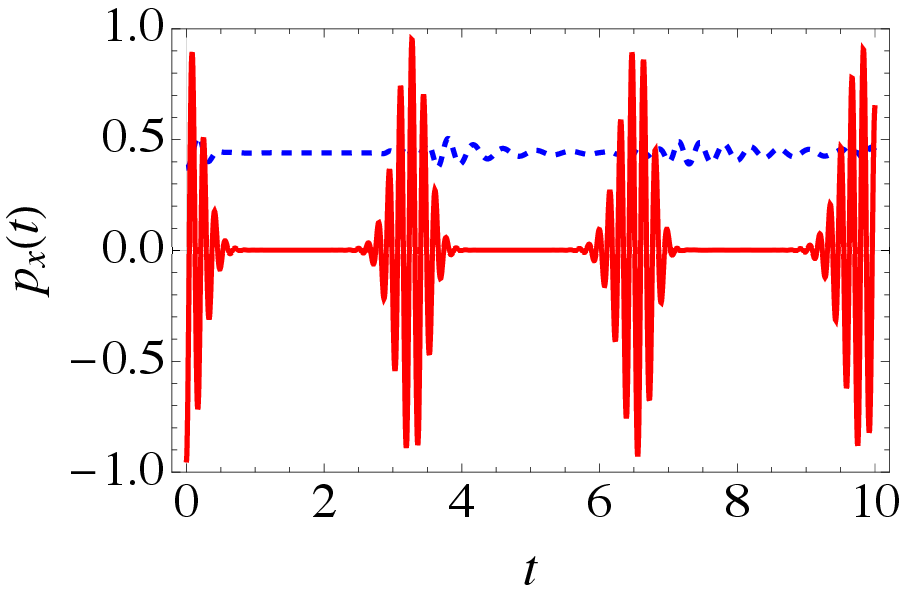}
 		\centering
		\caption{(Color online) Same as Fig.~\ref{weakcoupling}, except that now we have  $g = 1$ and  $\Delta_0 = 10$.}
		\label{strongcoupling}
\end{figure}

\begin{figure}[t]
 		\includegraphics[scale = 0.8]{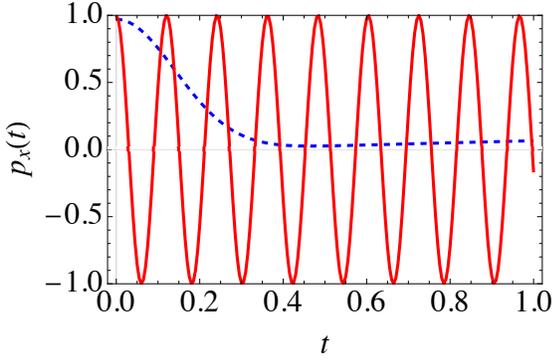}
 		\centering
		\caption{(Color online) Same as Fig.~\ref{weakcoupling}, except that now we have $\beta=10$, $g = 1$ and $\varepsilon_i = 0.01$. }
		\label{weakenv}
\end{figure}

We should also note that with a larger spin environment, the effect of the initial correlations is more pronounced. This is illustrated in Fig.~\ref{environment}; one can compare Fig.~\ref{environment}, where $N = 250$, with Fig.~\ref{weakcoupling} ($N = 50$) to see the effect of the increased number of environmental spins.  
We can also investigate how the tunneling amplitude of the central spin system affects the system dynamics. As shown in Fig.~\ref{Delta}, where we have increased the tunneling amplitude to $\Delta_0=10$, with coupling $g=0.05$, the dynamical difference is still evident. With the same tunneling amplitude, if the system-environment coupling strength is made even stronger, there is an even more significant difference [see Fig.~\ref{strongcoupling}]. The difference in the dynamics persists with different values of the energy bias of the environment as well [see Fig.~\ref{weakenv}].
\begin{figure}[t]
	\includegraphics[scale = 0.8]{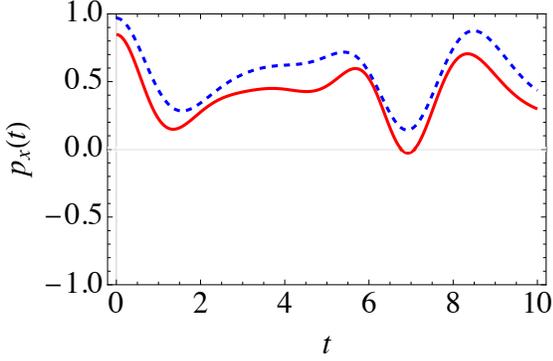}
	\centering	
	\caption{(Color online) Graph of $p_x(t)$ versus time $t$ for strong system-environment coupling without initial correlations (dashed, blue line) and with initial correlations (solid, red line). Here we have considered the interactions between the spins of environment $\kappa=0.1$, and we have set $\Delta_0= 1$. We also have $g_i = 0.5$, $\varepsilon_0=5, \varepsilon = 2$, $\varepsilon_i = 1$, $\beta = 1$, and $N = 10$.}
	\label{lowInterspin}
\end{figure}
Finally, let us consider the scenario where the environment spins are also interacting with each other. Once again, in general, we do see that the initial correlations play a significant role [see Fig.~\ref{lowInterspin}].

\begin{figure}[t]
	\includegraphics[scale = 0.8]{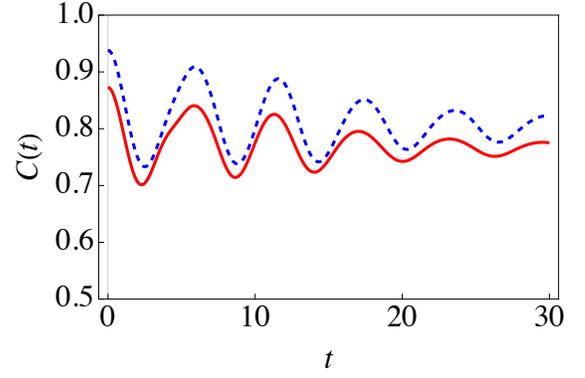}
	\centering
	\caption{(Color online) Plot of concurrence between the qubits versus time $t$ for relatively weak system-environment coupling strength $g_i = 0.05$ without initial correlations (dashed, blue line) and with initial correlations (solid, red line). We have also assumed that spins are not interacting with each other, that is, we have $\kappa = 0$. We have taken environment energy level spacing $\varepsilon_i=1$, and the other system-environment parameters are $\varepsilon^{(i)}_0=5, \varepsilon^{(i)} = 2$, $\Delta^{(i)}_{0} = 1$, $\beta =1$ and $N = 50$.}
	\label{2QweakCoupling}
\end{figure}

\begin{figure}[t]
    \includegraphics[scale = 0.8]{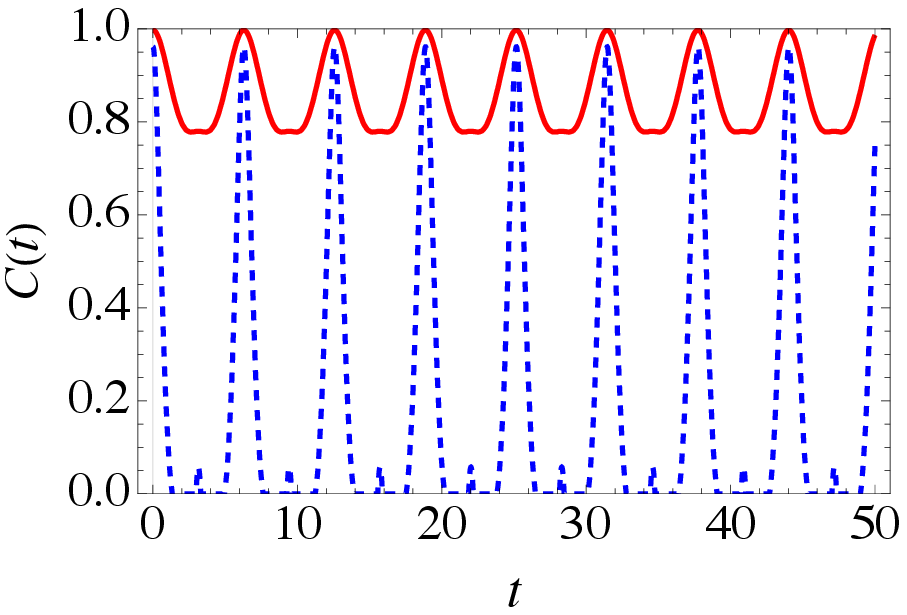}
 	\centering
 	\caption{(Color online) Same as Fig. \ref{2QweakCoupling}, except that now we have $\beta =3$ and $g_i =0.5$.}
 	\label{2Qlowtemp}
\end{figure}
\begin{figure}[t]
    \includegraphics[scale = 0.8]{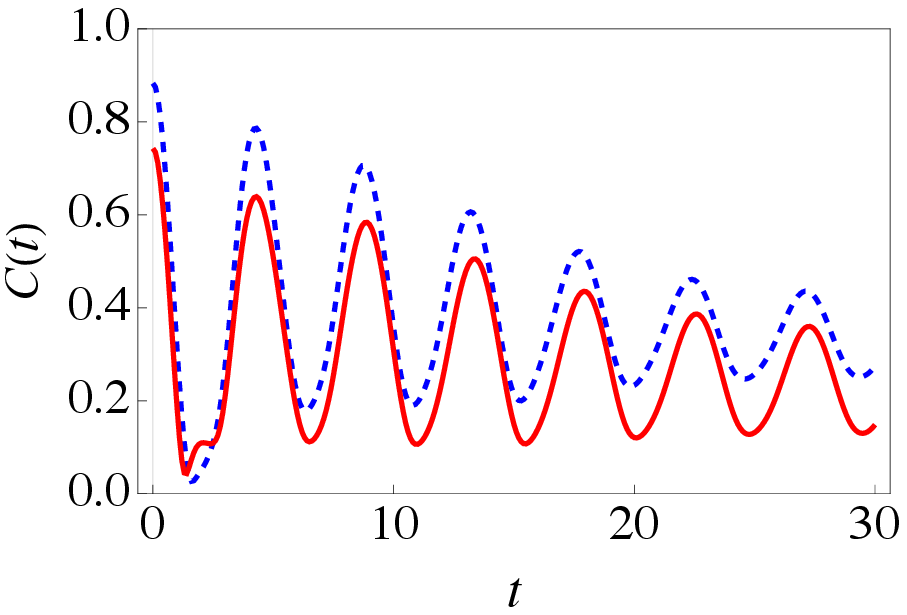}
 	\centering
 	\caption{(Color online) Same as Fig. \ref{2QweakCoupling}, expect that now we have $\kappa =0.5$.}
 	\label{qubitsinteraction}
\end{figure}

\section{Extension to two-qubit system}
\label{2qubits}
We now consider the case of two qubits interacting with the common spin environment. Again, our goal is to investigate the difference in dynamics for correlated and uncorrelated initial states. This could possibly reveal aspects of dynamics that may be absent in the single qubit case. An example is entanglement sudden death (ESD \cite{EberlyPRL2004,EberlyScience2007,EberlyScience2009}, where the entanglement between the two qubits vanishes in a very short time.

The total Hamiltonian is now
\begin{align}
    H_{\text{tot}} =
    \begin{cases}
      H^{(1)}_{\text{S0}} + H^{(2)}_{\text{S0}} + H_{\text{12}} + H^{(1)}_{\text{SE}} + H^{(2)}_{\text{SE}} + H_E & t\le 0,\\
      H^{(1)}_S + H^{(2)}_S + H_{\text{12}} + H^{(1)}_{\text{SE}} + H^{(2)}_{\text{SE}} + H_E & t > 0,\\
    \end{cases}   
\end{align}
with 
\begin{align}
    H^{(i)}_{\text{S0}}
    &= \frac{\varepsilon^{(i)}_0}{2} \sigma^{(i)}_{z} + \frac{\Delta^{(i)}_0}{2} \sigma^{(i)}_{x},
\\
	H^{(i)}_{\text{S}}
	&= \frac{\varepsilon^{(i)}}{2} \sigma^{(i)}_{z} + \frac{\Delta^{(i)}_0}{2} \sigma^{(i)}_{x},
\\
    H_{\text{12}}
    &= \kappa \sigma^{(1)}_{z}\sigma^{(2)}_{z},
\\
    H^{(1)}_{\text{SE}}
    &= \frac{1}{2} \sigma^{(1)}_{z} \otimes \sum_{i=1}^{N} g_i\sigma^{(i)}_z , 
\\
    H^{(2)}_{\text{SE}}
    &= \frac{1}{2} \sigma^{(2)}_{z} \otimes \sum_{i=1}^{N} g_i\sigma^{(i)}_z ,
\\
    H_E
    &=\sum_{i=1}^{N} \frac{\varepsilon_i}{2} \sigma^{(i)}_z + \sum_{i=1}^{N} \alpha_i \sigma^{(i)}_z \sigma^{(i+1)}_z,
\end{align}
with $i=1,2$. Here the qubits are labeled as 1 and 2 with $\varepsilon^{(1)}_0$ and $\varepsilon^{(2)}_0$ the energy bias terms and $\Delta^{(1)}_0$ and $\Delta^{(2)}_0$ the tunneling amplitudes of the central qubit 1 and qubit 2 respectively. Both qubits are coupled with each other by $H_{\text{12}}$. We aim to look at the dynamics of entanglement between the two-qubit system, starting from correlated and uncorrelated initial states. To begin, let us comment on the initial state preparation. We prepare our initial state such that, starting from the thermal equilibrium state, the two qubits become entangled with each other. Note that with $\varepsilon^{(i)}_0 \gg \Delta^{(i)}_0$, our system initial state is (approximately) both spins `down' along the $z\text{-}$axis. We now apply the unitary operator (at $t = 0$)
\begin{align*}
    \text{CZ}
    =e^{i \frac{\pi}{4}\left(\sigma^{(1)}_x + \sigma^{(2)}_x -\sigma_x \otimes \sigma_x\right)},
\end{align*}
only on the system of two qubits to generate entanglement between them. We then have the two different initial states
\begin{align*}
    \rho_{\text{entangled}}^{\text{woc}}
    &= \frac{1}{Z_{\text{woc}}} \text{CZ} e^{-\beta \left(H^{(1)}_{\text{S0}} + H^{(2)}_{\text{S0}} + H_{12} + H_E \right)}\text{CZ}^{\dagger},
\\    
    \rho_{\text{entangled}}^{\text{wc}}
    &= \frac{1}{Z_{\text{wc}}} \text{CZ} e^{-\beta \left(H^{(1)}_{\text{S0}} + H^{(2)}_{\text{S0}} + H_{12}  + H^{(1)}_{\text{SE}} + H^{(2)}_{\text{SE}} + H_E \right)}\text{CZ}^{\dagger}.
\end{align*}
Here $Z_{\text{woc}}$ and $Z_{\text{wc}}$ are the partition functions for the corresponding states.

For simplicity, we first consider $\kappa=0$ (the direct qubit-qubit interaction is zero) and analytically calculate the reduced density matrices, starting from these two different initial states. To do so, we need to find the time evolution operator. The calculation is very similar to the single qubit case; therefore, we simply summarize the results. For the simple product initial state, we obtain 
\begin{align}
    &\rho^{\text{woc}}_S(t)&=\frac{1}{Z_{\text{woc}}}\sum_{n}k_n U^{(1)}_n(t) U^{(2)}_n(t)\rho^{\text{woc}}_SU^{(2)\dagger}_n(t) U^{(1)\dagger}_n(t),
\end{align}
where
\begin{align}
    U^{(i)}_n(t)
	=e^{-i\frac{\epsilon_n}{4}t}e^{-i\frac{\lambda_n}{2}t}\left\{\mathds{1}\cos(\widetilde{\Delta}^{(i)}_nt)-\frac{i\sin(\widetilde{\Delta}^{(i)}_nt)}{\widetilde{\Delta}^{(i)}_n}H^{(i)}_{S,n}\right\},
\end{align}
with $\widetilde{\Delta}^{(i)}_n = (1/2)\sqrt{(\varepsilon^{(i)}_{n})^2 + (\Delta^{(i)})^2} $, $H_{S,n}^{(i)} = \frac{\varepsilon_n^{(i)}}{2} \sigma_z^{(i)} - \frac{\Delta_0^{(i)}}{2}\sigma_x^{(i)}$, $\varepsilon^{(i)}_n = e_n + \varepsilon^{(i)}$, and $Z_{\text{woc}} = \sum_n k_n$. On the other hand, with the correlated initial state, we get
\begin{align}
    &\rho^{\text{wc}}_S(t)
    =\tr_E\left\{U(t)\varrho^{\text{wc}}_{\text{entangled}}U^\dagger(t)\right\},\nonumber
\\
    &=\frac{1}{Z_{\text{wc}}}\sum_{n}A_n k_n {U^{(1)}_n(t)U^{(2)}_n(t)\rho_{S}^{\text{wc}}U^{(1)\dagger}_n(t)U^{(2)\dagger}_n(t)},
\end{align}
where $Z_{\text{wc}} = \sum_n A_n k_n$ with $A_n = \Tr\left\{e^{-\beta \left(H^{(1)}_{\text {S0},n} + H^{(2)}_{\text {S0},n}\right)}\right\}$ appearing due to the effect of initial correlations. Note that $H_{\text{S0},n}^{(i)}$ is the same as $H_{S,n}^{(i)}$ except the change of energy bias ($\varepsilon_{0,n}^{(i)}$ belongs to $H_{\text{S0},n}^{(i)}$ and $\varepsilon_n^{(i)}$ belongs to $H_{\text{S},n}^{(i)}$).

Using our worked-out dynamics, we can look at the impact of the initial correlations on the entanglement dynamics. To quantify entanglement, we use the concurrence $C(t)$. The concurrence of a two-qubit state $\rho(t)$ is defined as $C(t) = \text{max}\left(0, \sqrt{\lambda_1}-\sqrt{\lambda_2}-\sqrt{\lambda_3}-\sqrt{\lambda_4} \right)$, where the $\lambda_i$ (with $i=1,2,3,4$) denote the eigenvalues in decreasing order of $\rho(t)\left(\sigma^{(1)}_{y} \otimes \sigma^{(2)}_{y} \right) \rho^*(t)\left(\sigma^{(1)}_{y} \otimes \sigma^{(2)}_{y} \right)$. The concurrence is one for a maximally entangled state and zero for unentangled, separable states. We first see what examine the weak coupling regime. Fig.~\ref{2QweakCoupling} shows that even at weak system-environment coupling strength, we have considerable differences in the dynamics, and this difference is even more apparent at lower temperatures [see Fig.~\ref{2Qlowtemp}]; in fact, the initial correlations can even be seen to significantly enhance the entanglement. 

For completeness, let us note that we can further investigate the dynamics by including the effect of the qubit-qubit interaction as well. That is, $\kappa$ is non-zero now. Following a similar formalism, the time evolution operator is now found to be
\begin{equation}
\label{lambdatotalunitarytimeoperator}
{U}(t)=\sum_{n=0}^{2^{N}-1}{U}_{n}^{(12)}(t)\ket{n}\bra{n},
\end{equation}
with
${U}_{n}^{(12)}(t)
={e^{-i\frac{\epsilon_{n}}{2}t}}{e^{-i\lambda_{n}t}}e^{-i(H_{S,n}^{(1)}+H_{S,n}^{(2)}+H_{12})t}$.
This operator helps us to write our final system state for both the correlated and uncorrelated cases as
\begin{align}
{\rho}^{\text{woc}}_S(t)
&=\frac{1}{Z_E}\sum_{n=0}^{2^{N}-1} k_n U^{(12)}_{n}(t){\rho}^{\text{wc}}_{\text{S0}}U_n^{(12)\dagger}(t),
\\
{\rho}^{\text{woc}}_S(t)
&=\frac{1}{Z_{\text{tot}}}\sum_{n=0}^{2^{N}-1}k_n A_n {U^{(12)}_{n}(t)}{\rho}^{\text{woc}}_{\text{S0}}{U^{(12)}_{n}}^{\dagger}(t).
\end{align}
The key difference is that now $A_n = \Tr\left\{e^{-\beta \left(H^{(1)}_{\text {S0},n} + H^{(2)}_{\text {S0},n} + H_{12}\right)}\right\}$. We illustrate the entanglement dynamics with $\kappa = 0.5$ in Fig.~\ref{qubitsinteraction}. Once again, the effect of the initial correlations is quite apparent. 

\section{Conclusion}

In conclusion, we have explored the dynamics of a central two-level system interacting with a spin environment, taking into account the system-environment correlations. In our model, both the diagonal and off-diagonal elements of the density matrix of the central system evolve. We found that the effects of the initial system-environment correlations generally cause a minimal difference in the regime of weak system-environment coupling and high temperatures. However, this difference becomes very significant when the system-environment coupling becomes stronger and the temperature is reduced. We then extended our results to two spins interacting with a common spin environment, thereby showing that entanglement dynamics are also affected by the initial correlations. Our results highlight the importance of taking into account the effect of the initial system-environment correlations, especially in the strong system-environment coupling regime.

\section*{Acknowledgments}
We acknowledge useful discussions with Mehwish Majeed.


\begin{thebibliography}{38}%
\makeatletter
\providecommand \@ifxundefined [1]{%
 \@ifx{#1\undefined}
}%
\providecommand \@ifnum [1]{%
 \ifnum #1\expandafter \@firstoftwo
 \else \expandafter \@secondoftwo
 \fi
}%
\providecommand \@ifx [1]{%
 \ifx #1\expandafter \@firstoftwo
 \else \expandafter \@secondoftwo
 \fi
}%
\providecommand \natexlab [1]{#1}%
\providecommand \enquote  [1]{``#1''}%
\providecommand \bibnamefont  [1]{#1}%
\providecommand \bibfnamefont [1]{#1}%
\providecommand \citenamefont [1]{#1}%
\providecommand \href@noop [0]{\@secondoftwo}%
\providecommand \href [0]{\begingroup \@sanitize@url \@href}%
\providecommand \@href[1]{\@@startlink{#1}\@@href}%
\providecommand \@@href[1]{\endgroup#1\@@endlink}%
\providecommand \@sanitize@url [0]{\catcode `\\12\catcode `\$12\catcode
  `\&12\catcode `\#12\catcode `\^12\catcode `\_12\catcode `\%12\relax}%
\providecommand \@@startlink[1]{}%
\providecommand \@@endlink[0]{}%
\providecommand \url  [0]{\begingroup\@sanitize@url \@url }%
\providecommand \@url [1]{\endgroup\@href {#1}{\urlprefix }}%
\providecommand \urlprefix  [0]{URL }%
\providecommand \Eprint [0]{\href }%
\providecommand \doibase [0]{https://doi.org/}%
\providecommand \selectlanguage [0]{\@gobble}%
\providecommand \bibinfo  [0]{\@secondoftwo}%
\providecommand \bibfield  [0]{\@secondoftwo}%
\providecommand \translation [1]{[#1]}%
\providecommand \BibitemOpen [0]{}%
\providecommand \bibitemStop [0]{}%
\providecommand \bibitemNoStop [0]{.\EOS\space}%
\providecommand \EOS [0]{\spacefactor3000\relax}%
\providecommand \BibitemShut  [1]{\csname bibitem#1\endcsname}%
\let\auto@bib@innerbib\@empty
\bibitem [{\citenamefont {Breuer}\ and\ \citenamefont
  {Petruccione}(2007)}]{BPbook}%
  \BibitemOpen
  \bibfield  {author} {\bibinfo {author} {\bibfnamefont {H.-P.}\ \bibnamefont
  {Breuer}}\ and\ \bibinfo {author} {\bibfnamefont {F.}~\bibnamefont
  {Petruccione}},\ }\href@noop {} {\emph {\bibinfo {title} {The Theory of Open
  Quantum Systems}}}\ (\bibinfo  {publisher} {Oxford University Press},\
  \bibinfo {address} {Oxford},\ \bibinfo {year} {2007})\BibitemShut {NoStop}%
\bibitem [{\citenamefont {Weiss}(2008)}]{Weissbook}%
  \BibitemOpen
  \bibfield  {author} {\bibinfo {author} {\bibfnamefont {U.}~\bibnamefont
  {Weiss}},\ }\href@noop {} {\emph {\bibinfo {title} {Quantum dissipative
  systems}}}\ (\bibinfo  {publisher} {World Scientific},\ \bibinfo {address}
  {Singapore},\ \bibinfo {year} {2008})\BibitemShut {NoStop}%
\bibitem [{\citenamefont {Modi}(2011)}]{Modi2011}%
  \BibitemOpen
  \bibfield  {author} {\bibinfo {author} {\bibfnamefont {K.}~\bibnamefont
  {Modi}},\ }\bibfield  {title} {\bibinfo {title} {Preparation of states in
  open quantum mechanics},\ }\href {https://doi.org/10.1142/S1230161211000170}
  {\bibfield  {journal} {\bibinfo  {journal} {Open Syst. Inf. Dyn.}\ }\textbf
  {\bibinfo {volume} {18}},\ \bibinfo {pages} {253} (\bibinfo {year}
  {2011})}\BibitemShut {NoStop}%
\bibitem [{\citenamefont {Hakim}\ and\ \citenamefont
  {Ambegaokar}(1985)}]{HakimPRA1985}%
  \BibitemOpen
  \bibfield  {author} {\bibinfo {author} {\bibfnamefont {V.}~\bibnamefont
  {Hakim}}\ and\ \bibinfo {author} {\bibfnamefont {V.}~\bibnamefont
  {Ambegaokar}},\ }\bibfield  {title} {\bibinfo {title} {Quantum theory of a
  free particle interacting with a linearly dissipative environment},\ }\href
  {https://doi.org/10.1103/PhysRevA.32.423} {\bibfield  {journal} {\bibinfo
  {journal} {Phys. Rev. A}\ }\textbf {\bibinfo {volume} {32}},\ \bibinfo
  {pages} {423} (\bibinfo {year} {1985})}\BibitemShut {NoStop}%
\bibitem [{\citenamefont {Haake}\ and\ \citenamefont
  {Reibold}(1985)}]{HaakePRA1985}%
  \BibitemOpen
  \bibfield  {author} {\bibinfo {author} {\bibfnamefont {F.}~\bibnamefont
  {Haake}}\ and\ \bibinfo {author} {\bibfnamefont {R.}~\bibnamefont
  {Reibold}},\ }\bibfield  {title} {\bibinfo {title} {Strong damping and
  low-temperature anomalies for the harmonic oscillator},\ }\href
  {https://doi.org/10.1103/PhysRevA.32.2462} {\bibfield  {journal} {\bibinfo
  {journal} {Phys. Rev. A}\ }\textbf {\bibinfo {volume} {32}},\ \bibinfo
  {pages} {2462} (\bibinfo {year} {1985})}\BibitemShut {NoStop}%
\bibitem [{\citenamefont {Grabert}\ \emph {et~al.}(1988)\citenamefont
  {Grabert}, \citenamefont {Schramm},\ and\ \citenamefont
  {Ingold}}]{Grabert1988}%
  \BibitemOpen
  \bibfield  {author} {\bibinfo {author} {\bibfnamefont {H.}~\bibnamefont
  {Grabert}}, \bibinfo {author} {\bibfnamefont {P.}~\bibnamefont {Schramm}},\
  and\ \bibinfo {author} {\bibfnamefont {G.-L.}\ \bibnamefont {Ingold}},\
  }\bibfield  {title} {\bibinfo {title} {Quantum brownian motion: The
  functional integral approach},\ }\href
  {https://doi.org/10.1016/0370-1573(88)90023-3} {\bibfield  {journal}
  {\bibinfo  {journal} {Phys. Rep.}\ }\textbf {\bibinfo {volume} {168}},\
  \bibinfo {pages} {115} (\bibinfo {year} {1988})}\BibitemShut {NoStop}%
\bibitem [{\citenamefont {Smith}\ and\ \citenamefont
  {Caldeira}(1990)}]{SmithPRA1990}%
  \BibitemOpen
  \bibfield  {author} {\bibinfo {author} {\bibfnamefont {C.~M.}\ \bibnamefont
  {Smith}}\ and\ \bibinfo {author} {\bibfnamefont {A.~O.}\ \bibnamefont
  {Caldeira}},\ }\bibfield  {title} {\bibinfo {title} {Application of the
  generalized feynman-vernon approach to a simple system: The damped harmonic
  oscillator},\ }\href {https://doi.org/10.1103/PhysRevA.41.3103} {\bibfield
  {journal} {\bibinfo  {journal} {Phys. Rev. A}\ }\textbf {\bibinfo {volume}
  {41}},\ \bibinfo {pages} {3103} (\bibinfo {year} {1990})}\BibitemShut
  {NoStop}%
\bibitem [{\citenamefont {Karrlein}\ and\ \citenamefont
  {Grabert}(1997)}]{GrabertPRE1997}%
  \BibitemOpen
  \bibfield  {author} {\bibinfo {author} {\bibfnamefont {R.}~\bibnamefont
  {Karrlein}}\ and\ \bibinfo {author} {\bibfnamefont {H.}~\bibnamefont
  {Grabert}},\ }\bibfield  {title} {\bibinfo {title} {Exact time evolution and
  master equations for the damped harmonic oscillator},\ }\href
  {https://doi.org/10.1103/PhysRevE.55.153} {\bibfield  {journal} {\bibinfo
  {journal} {Phys. Rev. E}\ }\textbf {\bibinfo {volume} {55}},\ \bibinfo
  {pages} {153} (\bibinfo {year} {1997})}\BibitemShut {NoStop}%
\bibitem [{\citenamefont {D\'avila~Romero}\ and\ \citenamefont
  {Pablo~Paz}(1997)}]{PazPRA1997}%
  \BibitemOpen
  \bibfield  {author} {\bibinfo {author} {\bibfnamefont {L.}~\bibnamefont
  {D\'avila~Romero}}\ and\ \bibinfo {author} {\bibfnamefont {J.}~\bibnamefont
  {Pablo~Paz}},\ }\bibfield  {title} {\bibinfo {title} {Decoherence and initial
  correlations in quantum brownian motion},\ }\href
  {https://doi.org/10.1103/PhysRevA.55.4070} {\bibfield  {journal} {\bibinfo
  {journal} {Phys. Rev. A}\ }\textbf {\bibinfo {volume} {55}},\ \bibinfo
  {pages} {4070} (\bibinfo {year} {1997})}\BibitemShut {NoStop}%
\bibitem [{\citenamefont {Lutz}(2003)}]{LutzPRA2003}%
  \BibitemOpen
  \bibfield  {author} {\bibinfo {author} {\bibfnamefont {E.}~\bibnamefont
  {Lutz}},\ }\bibfield  {title} {\bibinfo {title} {Effect of initial
  correlations on short-time decoherence},\ }\href
  {https://doi.org/10.1103/PhysRevA.67.022109} {\bibfield  {journal} {\bibinfo
  {journal} {Phys. Rev. A}\ }\textbf {\bibinfo {volume} {67}},\ \bibinfo
  {pages} {022109} (\bibinfo {year} {2003})}\BibitemShut {NoStop}%
\bibitem [{\citenamefont {Banerjee}\ and\ \citenamefont
  {Ghosh}(2003)}]{BanerjeePRE2003}%
  \BibitemOpen
  \bibfield  {author} {\bibinfo {author} {\bibfnamefont {S.}~\bibnamefont
  {Banerjee}}\ and\ \bibinfo {author} {\bibfnamefont {R.}~\bibnamefont
  {Ghosh}},\ }\bibfield  {title} {\bibinfo {title} {General quantum brownian
  motion with initially correlated and nonlinearly coupled environment},\
  }\href {https://doi.org/10.1103/PhysRevE.67.056120} {\bibfield  {journal}
  {\bibinfo  {journal} {Phys. Rev. E}\ }\textbf {\bibinfo {volume} {67}},\
  \bibinfo {pages} {056120} (\bibinfo {year} {2003})}\BibitemShut {NoStop}%
\bibitem [{\citenamefont {van Kampen}(2004)}]{vanKampen2004}%
  \BibitemOpen
  \bibfield  {author} {\bibinfo {author} {\bibfnamefont {N.~G.}\ \bibnamefont
  {van Kampen}},\ }\href {https://doi.org/10.1023/B:JOSS.0000022383.06086.6c}
  {\bibfield  {journal} {\bibinfo  {journal} {J. Stat. Phys.}\ }\textbf
  {\bibinfo {volume} {115}},\ \bibinfo {pages} {1057} (\bibinfo {year}
  {2004})}\BibitemShut {NoStop}%
\bibitem [{\citenamefont {Ban}(2009)}]{BanPRA2009}%
  \BibitemOpen
  \bibfield  {author} {\bibinfo {author} {\bibfnamefont {M.}~\bibnamefont
  {Ban}},\ }\bibfield  {title} {\bibinfo {title} {Quantum master equation for
  dephasing of a two-level system with an initial correlation},\ }\href
  {https://doi.org/10.1103/PhysRevA.80.064103} {\bibfield  {journal} {\bibinfo
  {journal} {Phys. Rev. A}\ }\textbf {\bibinfo {volume} {80}},\ \bibinfo
  {pages} {064103} (\bibinfo {year} {2009})}\BibitemShut {NoStop}%
\bibitem [{\citenamefont {Campisi}\ \emph {et~al.}(2009)\citenamefont
  {Campisi}, \citenamefont {Talkner},\ and\ \citenamefont
  {H\"anggi}}]{HanggiPRL2009}%
  \BibitemOpen
  \bibfield  {author} {\bibinfo {author} {\bibfnamefont {M.}~\bibnamefont
  {Campisi}}, \bibinfo {author} {\bibfnamefont {P.}~\bibnamefont {Talkner}},\
  and\ \bibinfo {author} {\bibfnamefont {P.}~\bibnamefont {H\"anggi}},\
  }\bibfield  {title} {\bibinfo {title} {Fluctuation theorem for arbitrary open
  quantum systems},\ }\href {https://doi.org/10.1103/PhysRevLett.102.210401}
  {\bibfield  {journal} {\bibinfo  {journal} {Phys. Rev. Lett.}\ }\textbf
  {\bibinfo {volume} {102}},\ \bibinfo {pages} {210401} (\bibinfo {year}
  {2009})}\BibitemShut {NoStop}%
\bibitem [{\citenamefont {Uchiyama}\ and\ \citenamefont
  {Aihara}(2010)}]{UchiyamaPRA2010}%
  \BibitemOpen
  \bibfield  {author} {\bibinfo {author} {\bibfnamefont {C.}~\bibnamefont
  {Uchiyama}}\ and\ \bibinfo {author} {\bibfnamefont {M.}~\bibnamefont
  {Aihara}},\ }\bibfield  {title} {\bibinfo {title} {Role of initial quantum
  correlation in transient linear response},\ }\href
  {https://doi.org/10.1103/PhysRevA.82.044104} {\bibfield  {journal} {\bibinfo
  {journal} {Phys. Rev. A}\ }\textbf {\bibinfo {volume} {82}},\ \bibinfo
  {pages} {044104} (\bibinfo {year} {2010})}\BibitemShut {NoStop}%
\bibitem [{\citenamefont {Dijkstra}\ and\ \citenamefont
  {Tanimura}(2010)}]{TanimuraPRL2010}%
  \BibitemOpen
  \bibfield  {author} {\bibinfo {author} {\bibfnamefont {A.~G.}\ \bibnamefont
  {Dijkstra}}\ and\ \bibinfo {author} {\bibfnamefont {Y.}~\bibnamefont
  {Tanimura}},\ }\bibfield  {title} {\bibinfo {title} {Non-markovian
  entanglement dynamics in the presence of system-bath coherence},\ }\href
  {https://doi.org/10.1103/PhysRevLett.104.250401} {\bibfield  {journal}
  {\bibinfo  {journal} {Phys. Rev. Lett.}\ }\textbf {\bibinfo {volume} {104}},\
  \bibinfo {pages} {250401} (\bibinfo {year} {2010})}\BibitemShut {NoStop}%
\bibitem [{\citenamefont {Smirne}\ \emph {et~al.}(2010)\citenamefont {Smirne},
  \citenamefont {Breuer}, \citenamefont {Piilo},\ and\ \citenamefont
  {Vacchini}}]{SmirnePRA2010}%
  \BibitemOpen
  \bibfield  {author} {\bibinfo {author} {\bibfnamefont {A.}~\bibnamefont
  {Smirne}}, \bibinfo {author} {\bibfnamefont {H.-P.}\ \bibnamefont {Breuer}},
  \bibinfo {author} {\bibfnamefont {J.}~\bibnamefont {Piilo}},\ and\ \bibinfo
  {author} {\bibfnamefont {B.}~\bibnamefont {Vacchini}},\ }\bibfield  {title}
  {\bibinfo {title} {Initial correlations in open-systems dynamics: The
  jaynes-cummings model},\ }\href {https://doi.org/10.1103/PhysRevA.82.062114}
  {\bibfield  {journal} {\bibinfo  {journal} {Phys. Rev. A}\ }\textbf {\bibinfo
  {volume} {82}},\ \bibinfo {pages} {062114} (\bibinfo {year}
  {2010})}\BibitemShut {NoStop}%
\bibitem [{\citenamefont {Dajka}\ and\ \citenamefont
  {\L{}uczka}(2010)}]{DajkaPRA2010}%
  \BibitemOpen
  \bibfield  {author} {\bibinfo {author} {\bibfnamefont {J.}~\bibnamefont
  {Dajka}}\ and\ \bibinfo {author} {\bibfnamefont {J.}~\bibnamefont
  {\L{}uczka}},\ }\bibfield  {title} {\bibinfo {title} {Distance growth of
  quantum states due to initial system-environment correlations},\ }\href
  {https://doi.org/10.1103/PhysRevA.82.012341} {\bibfield  {journal} {\bibinfo
  {journal} {Phys. Rev. A}\ }\textbf {\bibinfo {volume} {82}},\ \bibinfo
  {pages} {012341} (\bibinfo {year} {2010})}\BibitemShut {NoStop}%
\bibitem [{\citenamefont {Zhang}\ \emph {et~al.}(2010)\citenamefont {Zhang},
  \citenamefont {Zou}, \citenamefont {Xia},\ and\ \citenamefont
  {Guo}}]{ZhangPRA2010}%
  \BibitemOpen
  \bibfield  {author} {\bibinfo {author} {\bibfnamefont {Y.-J.}\ \bibnamefont
  {Zhang}}, \bibinfo {author} {\bibfnamefont {X.-B.}\ \bibnamefont {Zou}},
  \bibinfo {author} {\bibfnamefont {Y.-J.}\ \bibnamefont {Xia}},\ and\ \bibinfo
  {author} {\bibfnamefont {G.-C.}\ \bibnamefont {Guo}},\ }\bibfield  {title}
  {\bibinfo {title} {Different entanglement dynamical behaviors due to initial
  system-environment correlations},\ }\href
  {https://doi.org/10.1103/PhysRevA.82.022108} {\bibfield  {journal} {\bibinfo
  {journal} {Phys. Rev. A}\ }\textbf {\bibinfo {volume} {82}},\ \bibinfo
  {pages} {022108} (\bibinfo {year} {2010})}\BibitemShut {NoStop}%
\bibitem [{\citenamefont {Tan}\ and\ \citenamefont {Zhang}(2011)}]{TanPRA2011}%
  \BibitemOpen
  \bibfield  {author} {\bibinfo {author} {\bibfnamefont {H.-T.}\ \bibnamefont
  {Tan}}\ and\ \bibinfo {author} {\bibfnamefont {W.-M.}\ \bibnamefont
  {Zhang}},\ }\bibfield  {title} {\bibinfo {title} {Non-markovian dynamics of
  an open quantum system with initial system-reservoir correlations: A
  nanocavity coupled to a coupled-resonator optical waveguide},\ }\href
  {https://doi.org/10.1103/PhysRevA.83.032102} {\bibfield  {journal} {\bibinfo
  {journal} {Phys. Rev. A}\ }\textbf {\bibinfo {volume} {83}},\ \bibinfo
  {pages} {032102} (\bibinfo {year} {2011})}\BibitemShut {NoStop}%
\bibitem [{\citenamefont {Lee}\ \emph {et~al.}(2012)\citenamefont {Lee},
  \citenamefont {Cao},\ and\ \citenamefont {Gong}}]{CKLeePRE2012}%
  \BibitemOpen
  \bibfield  {author} {\bibinfo {author} {\bibfnamefont {C.~K.}\ \bibnamefont
  {Lee}}, \bibinfo {author} {\bibfnamefont {J.}~\bibnamefont {Cao}},\ and\
  \bibinfo {author} {\bibfnamefont {J.}~\bibnamefont {Gong}},\ }\bibfield
  {title} {\bibinfo {title} {Noncanonical statistics of a spin-boson model:
  Theory and exact monte carlo simulations},\ }\href
  {https://doi.org/10.1103/PhysRevE.86.021109} {\bibfield  {journal} {\bibinfo
  {journal} {Phys. Rev. E}\ }\textbf {\bibinfo {volume} {86}},\ \bibinfo
  {pages} {021109} (\bibinfo {year} {2012})}\BibitemShut {NoStop}%
\bibitem [{\citenamefont {Morozov}\ \emph {et~al.}(2012)\citenamefont
  {Morozov}, \citenamefont {Mathey},\ and\ \citenamefont
  {R\"opke}}]{MorozovPRA2012}%
  \BibitemOpen
  \bibfield  {author} {\bibinfo {author} {\bibfnamefont {V.~G.}\ \bibnamefont
  {Morozov}}, \bibinfo {author} {\bibfnamefont {S.}~\bibnamefont {Mathey}},\
  and\ \bibinfo {author} {\bibfnamefont {G.}~\bibnamefont {R\"opke}},\
  }\bibfield  {title} {\bibinfo {title} {Decoherence in an exactly solvable
  qubit model with initial qubit-environment correlations},\ }\href
  {https://doi.org/10.1103/PhysRevA.85.022101} {\bibfield  {journal} {\bibinfo
  {journal} {Phys. Rev. A}\ }\textbf {\bibinfo {volume} {85}},\ \bibinfo
  {pages} {022101} (\bibinfo {year} {2012})}\BibitemShut {NoStop}%
\bibitem [{\citenamefont {Semin}\ \emph {et~al.}(2012)\citenamefont {Semin},
  \citenamefont {Sinayskiy},\ and\ \citenamefont {Petruccione}}]{SeminPRA2012}%
  \BibitemOpen
  \bibfield  {author} {\bibinfo {author} {\bibfnamefont {V.}~\bibnamefont
  {Semin}}, \bibinfo {author} {\bibfnamefont {I.}~\bibnamefont {Sinayskiy}},\
  and\ \bibinfo {author} {\bibfnamefont {F.}~\bibnamefont {Petruccione}},\
  }\bibfield  {title} {\bibinfo {title} {Initial correlation in a system of a
  spin coupled to a spin bath through an intermediate spin},\ }\href
  {https://doi.org/10.1103/PhysRevA.86.062114} {\bibfield  {journal} {\bibinfo
  {journal} {Phys. Rev. A}\ }\textbf {\bibinfo {volume} {86}},\ \bibinfo
  {pages} {062114} (\bibinfo {year} {2012})}\BibitemShut {NoStop}%
\bibitem [{\citenamefont {Chaudhry}\ and\ \citenamefont
  {Gong}(2013{\natexlab{a}})}]{ChaudhryPRA2013a}%
  \BibitemOpen
  \bibfield  {author} {\bibinfo {author} {\bibfnamefont {A.~Z.}\ \bibnamefont
  {Chaudhry}}\ and\ \bibinfo {author} {\bibfnamefont {J.}~\bibnamefont
  {Gong}},\ }\bibfield  {title} {\bibinfo {title} {Amplification and
  suppression of system-bath-correlation effects in an open many-body system},\
  }\href {https://doi.org/10.1103/PhysRevA.87.012129} {\bibfield  {journal}
  {\bibinfo  {journal} {Phys. Rev. A}\ }\textbf {\bibinfo {volume} {87}},\
  \bibinfo {pages} {012129} (\bibinfo {year} {2013}{\natexlab{a}})}\BibitemShut
  {NoStop}%
\bibitem [{\citenamefont {Chaudhry}\ and\ \citenamefont
  {Gong}(2013{\natexlab{b}})}]{ChaudhryPRA2013b}%
  \BibitemOpen
  \bibfield  {author} {\bibinfo {author} {\bibfnamefont {A.~Z.}\ \bibnamefont
  {Chaudhry}}\ and\ \bibinfo {author} {\bibfnamefont {J.}~\bibnamefont
  {Gong}},\ }\bibfield  {title} {\bibinfo {title} {Role of initial
  system-environment correlations: A master equation approach},\ }\href
  {https://doi.org/10.1103/PhysRevA.88.052107} {\bibfield  {journal} {\bibinfo
  {journal} {Phys. Rev. A}\ }\textbf {\bibinfo {volume} {88}},\ \bibinfo
  {pages} {052107} (\bibinfo {year} {2013}{\natexlab{b}})}\BibitemShut
  {NoStop}%
\bibitem [{\citenamefont {Chaudhry}\ and\ \citenamefont
  {Gong}(2013{\natexlab{c}})}]{ChaudhryCJC2013}%
  \BibitemOpen
  \bibfield  {author} {\bibinfo {author} {\bibfnamefont {A.~Z.}\ \bibnamefont
  {Chaudhry}}\ and\ \bibinfo {author} {\bibfnamefont {J.}~\bibnamefont
  {Gong}},\ }\bibfield  {title} {\bibinfo {title} {The effect of state
  preparation in a many-body system},\ }\href@noop {} {\bibfield  {journal}
  {\bibinfo  {journal} {Can. J. Chem.}\ }\textbf {\bibinfo {volume} {92}},\
  \bibinfo {pages} {119} (\bibinfo {year} {2013}{\natexlab{c}})}\BibitemShut
  {NoStop}%
\bibitem [{\citenamefont {Reina}\ \emph {et~al.}(2014)\citenamefont {Reina},
  \citenamefont {Susa},\ and\ \citenamefont {Fanchini}}]{FanchiniSciRep2014}%
  \BibitemOpen
  \bibfield  {author} {\bibinfo {author} {\bibfnamefont {J.}~\bibnamefont
  {Reina}}, \bibinfo {author} {\bibfnamefont {C.}~\bibnamefont {Susa}},\ and\
  \bibinfo {author} {\bibfnamefont {F.}~\bibnamefont {Fanchini}},\ }\bibfield
  {title} {\bibinfo {title} {Extracting information from qubit-environment
  correlations},\ }\href {https://doi.org/10.1038/srep07443} {\bibfield
  {journal} {\bibinfo  {journal} {Sci. Rep.}\ }\textbf {\bibinfo {volume}
  {4}},\ \bibinfo {pages} {7443} (\bibinfo {year} {2014})}\BibitemShut
  {NoStop}%
\bibitem [{\citenamefont {Zhang}\ \emph {et~al.}(2015)\citenamefont {Zhang},
  \citenamefont {Han}, \citenamefont {Xia}, \citenamefont {Yu},\ and\
  \citenamefont {Fan}}]{FanSciRep2015}%
  \BibitemOpen
  \bibfield  {author} {\bibinfo {author} {\bibfnamefont {Y.-J.}\ \bibnamefont
  {Zhang}}, \bibinfo {author} {\bibfnamefont {W.}~\bibnamefont {Han}}, \bibinfo
  {author} {\bibfnamefont {Y.-J.}\ \bibnamefont {Xia}}, \bibinfo {author}
  {\bibfnamefont {Y.-M.}\ \bibnamefont {Yu}},\ and\ \bibinfo {author}
  {\bibfnamefont {H.}~\bibnamefont {Fan}},\ }\bibfield  {title} {\bibinfo
  {title} {Role of initial system-bath correlation on coherence trapping},\
  }\href {https://doi.org/10.1038/srep13359} {\bibfield  {journal} {\bibinfo
  {journal} {Sci. Rep.}\ }\textbf {\bibinfo {volume} {5}},\ \bibinfo {pages}
  {13359} (\bibinfo {year} {2015})}\BibitemShut {NoStop}%
\bibitem [{\citenamefont {Chen}\ and\ \citenamefont
  {Goan}(2016)}]{ChenPRA2016}%
  \BibitemOpen
  \bibfield  {author} {\bibinfo {author} {\bibfnamefont {C.-C.}\ \bibnamefont
  {Chen}}\ and\ \bibinfo {author} {\bibfnamefont {H.-S.}\ \bibnamefont
  {Goan}},\ }\bibfield  {title} {\bibinfo {title} {Effects of initial
  system-environment correlations on open-quantum-system dynamics and state
  preparation},\ }\href {https://doi.org/10.1103/PhysRevA.93.032113} {\bibfield
   {journal} {\bibinfo  {journal} {Phys. Rev. A}\ }\textbf {\bibinfo {volume}
  {93}},\ \bibinfo {pages} {032113} (\bibinfo {year} {2016})}\BibitemShut
  {NoStop}%
\bibitem [{\citenamefont {de~Vega}\ and\ \citenamefont
  {Alonso}(2017)}]{VegaRMP2017}%
  \BibitemOpen
  \bibfield  {author} {\bibinfo {author} {\bibfnamefont {I.}~\bibnamefont
  {de~Vega}}\ and\ \bibinfo {author} {\bibfnamefont {D.}~\bibnamefont
  {Alonso}},\ }\bibfield  {title} {\bibinfo {title} {Dynamics of non-markovian
  open quantum systems},\ }\href {https://doi.org/10.1103/RevModPhys.89.015001}
  {\bibfield  {journal} {\bibinfo  {journal} {Rev. Mod. Phys.}\ }\textbf
  {\bibinfo {volume} {89}},\ \bibinfo {pages} {015001} (\bibinfo {year}
  {2017})}\BibitemShut {NoStop}%
\bibitem [{\citenamefont {Halimeh}\ and\ \citenamefont
  {de~Vega}(2017)}]{VegaPRA2017}%
  \BibitemOpen
  \bibfield  {author} {\bibinfo {author} {\bibfnamefont {J.~C.}\ \bibnamefont
  {Halimeh}}\ and\ \bibinfo {author} {\bibfnamefont {I.}~\bibnamefont
  {de~Vega}},\ }\bibfield  {title} {\bibinfo {title} {Weak-coupling master
  equation for arbitrary initial conditions},\ }\href
  {https://doi.org/10.1103/PhysRevA.95.052108} {\bibfield  {journal} {\bibinfo
  {journal} {Phys. Rev. A}\ }\textbf {\bibinfo {volume} {95}},\ \bibinfo
  {pages} {052108} (\bibinfo {year} {2017})}\BibitemShut {NoStop}%
\bibitem [{\citenamefont {Kitajima}\ \emph {et~al.}(2017)\citenamefont
  {Kitajima}, \citenamefont {Ban},\ and\ \citenamefont
  {Shibata}}]{ShibataJPhysA2017}%
  \BibitemOpen
  \bibfield  {author} {\bibinfo {author} {\bibfnamefont {S.}~\bibnamefont
  {Kitajima}}, \bibinfo {author} {\bibfnamefont {M.}~\bibnamefont {Ban}},\ and\
  \bibinfo {author} {\bibfnamefont {F.}~\bibnamefont {Shibata}},\ }\bibfield
  {title} {\bibinfo {title} {Expansion formulas for quantum master equations
  including initial correlation},\ }\href@noop {} {\bibfield  {journal}
  {\bibinfo  {journal} {J. Phys. A: Math. Theor}\ }\textbf {\bibinfo {volume}
  {50}},\ \bibinfo {pages} {125303} (\bibinfo {year} {2017})}\BibitemShut
  {NoStop}%
\bibitem [{\citenamefont {Buser}\ \emph {et~al.}(2017)\citenamefont {Buser},
  \citenamefont {Cerrillo}, \citenamefont {Schaller},\ and\ \citenamefont
  {Cao}}]{CaoPRA2017}%
  \BibitemOpen
  \bibfield  {author} {\bibinfo {author} {\bibfnamefont {M.}~\bibnamefont
  {Buser}}, \bibinfo {author} {\bibfnamefont {J.}~\bibnamefont {Cerrillo}},
  \bibinfo {author} {\bibfnamefont {G.}~\bibnamefont {Schaller}},\ and\
  \bibinfo {author} {\bibfnamefont {J.}~\bibnamefont {Cao}},\ }\bibfield
  {title} {\bibinfo {title} {Initial system-environment correlations via the
  transfer-tensor method},\ }\href {https://doi.org/10.1103/PhysRevA.96.062122}
  {\bibfield  {journal} {\bibinfo  {journal} {Phys. Rev. A}\ }\textbf {\bibinfo
  {volume} {96}},\ \bibinfo {pages} {062122} (\bibinfo {year}
  {2017})}\BibitemShut {NoStop}%
\bibitem [{\citenamefont {Cucchietti}\ \emph {et~al.}(2005)\citenamefont
  {Cucchietti}, \citenamefont {Paz},\ and\ \citenamefont
  {Zurek}}]{CucchiettiPRA2005}%
  \BibitemOpen
  \bibfield  {author} {\bibinfo {author} {\bibfnamefont {F.~M.}\ \bibnamefont
  {Cucchietti}}, \bibinfo {author} {\bibfnamefont {J.~P.}\ \bibnamefont
  {Paz}},\ and\ \bibinfo {author} {\bibfnamefont {W.~H.}\ \bibnamefont
  {Zurek}},\ }\bibfield  {title} {\bibinfo {title} {Decoherence from spin
  environments},\ }\href {https://doi.org/10.1103/PhysRevA.72.052113}
  {\bibfield  {journal} {\bibinfo  {journal} {Phys. Rev. A}\ }\textbf {\bibinfo
  {volume} {72}},\ \bibinfo {pages} {052113} (\bibinfo {year}
  {2005})}\BibitemShut {NoStop}%
\bibitem [{\citenamefont {Majeed}\ and\ \citenamefont
  {Chaudhry}(2019)}]{majeed2019effect}%
  \BibitemOpen
  \bibfield  {author} {\bibinfo {author} {\bibfnamefont {M.}~\bibnamefont
  {Majeed}}\ and\ \bibinfo {author} {\bibfnamefont {A.~Z.}\ \bibnamefont
  {Chaudhry}},\ }\bibfield  {title} {\bibinfo {title} {Effect of initial
  system--environment correlations with spin environments},\ }\href@noop {}
  {\bibfield  {journal} {\bibinfo  {journal} {The European Physical Journal D}\
  }\textbf {\bibinfo {volume} {73}},\ \bibinfo {pages} {1} (\bibinfo {year}
  {2019})}\BibitemShut {NoStop}%
\bibitem [{\citenamefont {Yu}\ and\ \citenamefont
  {Eberly}(2004)}]{EberlyPRL2004}%
  \BibitemOpen
  \bibfield  {author} {\bibinfo {author} {\bibfnamefont {T.}~\bibnamefont
  {Yu}}\ and\ \bibinfo {author} {\bibfnamefont {J.~H.}\ \bibnamefont
  {Eberly}},\ }\bibfield  {title} {\bibinfo {title} {Finite-time
  disentanglement via spontaneous emission},\ }\href
  {https://doi.org/10.1103/PhysRevLett.93.140404} {\bibfield  {journal}
  {\bibinfo  {journal} {Phys. Rev. Lett.}\ }\textbf {\bibinfo {volume} {93}},\
  \bibinfo {pages} {140404} (\bibinfo {year} {2004})}\BibitemShut {NoStop}%
\bibitem [{\citenamefont {Eberly}\ and\ \citenamefont
  {Yu}(2007)}]{EberlyScience2007}%
  \BibitemOpen
  \bibfield  {author} {\bibinfo {author} {\bibfnamefont {J.}~\bibnamefont
  {Eberly}}\ and\ \bibinfo {author} {\bibfnamefont {T.}~\bibnamefont {Yu}},\
  }\bibfield  {title} {\bibinfo {title} {The end of an entanglement},\
  }\href@noop {} {\bibfield  {journal} {\bibinfo  {journal} {Science}\ }\textbf
  {\bibinfo {volume} {316}},\ \bibinfo {pages} {555} (\bibinfo {year}
  {2007})}\BibitemShut {NoStop}%
\bibitem [{\citenamefont {Yu}\ and\ \citenamefont
  {Eberly}(2009)}]{EberlyScience2009}%
  \BibitemOpen
  \bibfield  {author} {\bibinfo {author} {\bibfnamefont {T.}~\bibnamefont
  {Yu}}\ and\ \bibinfo {author} {\bibfnamefont {J.~H.}\ \bibnamefont
  {Eberly}},\ }\bibfield  {title} {\bibinfo {title} {Sudden death of
  entanglement},\ }\href {https://doi.org/10.1126/science.1167343} {\bibfield
  {journal} {\bibinfo  {journal} {Science}\ }\textbf {\bibinfo {volume}
  {323}},\ \bibinfo {pages} {598} (\bibinfo {year} {2009})}\BibitemShut
  {NoStop}%
\end{thebibliography}
\end{document}